\documentclass[amsmat,amssymb,amsfonts,aps,prb,twocolumn,showpacs]{revtex4}

\usepackage{graphicx}
\usepackage{dcolumn}
\usepackage{bm}
\begin{document}

\title{Coherent transport in graphene nanoconstrictions}

\author{F. Mu\~noz-Rojas, D. Jacob,  J.
Fern\'andez-Rossier\footnote{jfrossier@ua.es} , and J. J.
Palacios }
\affiliation{Departamento de F\'{\i}sica Aplicada, Universidad de Alicante, San
Vicente del Raspeig, SPAIN}

\date{\today} 

\begin{abstract} 

We study the effect of a structural nanoconstriction on the  coherent transport
properties of  otherwise ideal zig-zag-edged infinitely long graphene ribbons.
The electronic structure is calculated with the  standard   one-orbital
tight-binding model and the linear conductance  is obtained using the Landauer
formula.   We find that, since the zero-bias current is carried in the bulk of
the ribbon, this is very robust with respect to a variety of constriction
geometries and edge defects. In contrast, the curve of zero-bias conductance
versus gate voltage departs from the $(2n+1) e^2/h$ staircase of the ideal case
as soon as a single atom is removed from the  sample.  We also find that
wedge-shaped constrictions can present non-conducting states fully localized in
the constriction close to the Fermi energy. The interest of these localized
states in regards the formation of quantum dots in graphene is discussed.

\end{abstract}

\maketitle

\section{Introduction}
The  recent fabrication of field effect transistor devices based both upon 
quasi-2D graphite quantum dots \cite{Bunch05} and upon  
 graphene \cite{Geim05,Kim05,Science06}
 (a single atomic layer of graphite), and the observation of a new type of
quantum Hall effect in the latter have triggered a huge interest in the
electronic properties of this system. Most of the results of standard
mesoscopic physics need to be revisited\cite{Katsnelson06,Beenakker06,Diffusive}
in the case of graphene because its electronic structure  is fundamentally
different from that of  metals and semiconductors where either a large density
of states (DOS) at the Fermi energy or a large gap determine the properties of
the materials.  Graphene is a semimetal with zero DOS at the Fermi energy and
zero  gap.  On top of that, the electronic structure close to the Fermi energy
has a conical shape with perfect electron-hole symmetry, identical to that of
two dimensional massless Dirac fermions\cite{Semenoff84}.

Here we consider graphene-based  one dimensional flat structures, the so called
graphene  nano-ribbons \cite{Fujita96,Wakabayashi99,Ezawa06,Brey06}.
As opposed to two dimensional 
graphene,   ribbons  can present a finite density of states at the Fermi energy which
dominates electrical transport in undoped or weakly doped samples.
Ideal graphene ribbons can be
considered as the  flat parent structures of carbon nanotubes, whose electronic
properties have been thoroughly studied\cite{Book}. Electronic transport in
carbon nanotubes has been studied in different regimes, including  ballistic
\cite{NT-Dai}, Coulomb Blockade\cite{Jarillo1} and Kondo \cite{KondoNT,
Jarillo2}. Progress in the fabrication of graphene based  nano-structures that
permits to study  transport in graphene ribbons  motivates this work.

The presence of edges makes the
electronic structure of graphene ribbons  different from that of 
nanotubes. Two types of idealized  edges  are usually considered: Armchair and
zig-zag \cite{Fujita96}.
Interestingly,   all the zig-zag and some of the armchair edges
result in  a band at the Fermi energy\cite{Fujita96}. In the case of
narrow zig-zag-edged ribbons, the top and bottom edge states can be
sufficiently close as to make hybridization possible, resulting  in two low
energy dispersive bands, symmetrically placed around the Fermi energy, $E=0$ .  
Depending on their width, nano-ribbons with armchair edges
can be metallic or insulating.
The different density of states for armchair and zig-zag edges has been
experimentally observed in Scanning Tunnel Microscope (STM) 
experiments with atomic resolution \cite{Niimi05,Kobayashi05}.

Coherent or quantum transport in graphene ribbons has been studied previously,
both for ideal \cite{PRBGuinea06} and defective
\cite{Sigrist2000,Wakabayashi2001} cases in the infinite length $L$ case.
The case of disorder free graphene ribbons with  finite $L$ comparable to the
width $W$ has also been studied in connection with the experimental
measurement of a minimum conductivity in mesoscopic size graphene layers
\cite{Katsnelson06,Beenakker06}. 
In the spirit of quantum point contact physics, 
in this work we study the coherent transport of infinitely long
 graphene ribbons, narrower than  $W\simeq5$nm,  with a
structural nanometric constriction like the one schematically shown in 
Fig. (\ref{fig:3:LDR}).  
This type of structure can also be considered as  an idealization of
an all-carbon single-molecule junction. In conventional single-molecule
junctions the organic molecule is attached to metallic electrodes. 
Here these are
replaced by perfect graphene ribbons and the role of the molecule is played by a
geometrical constriction.
Due to the very different electronic structure 
displayed by different all-carbon nano-structures\cite{Book} the conduction properties of 
these systems do not appear obvious {\em a priori} to us.

We calculate the electronic structure in the one orbital
tight binding approximation. The relevant orbital is the $p_z$,
since the sp$^2$ orbitals form bonding and
anti-bonding  states very far away in energy.
In the TB approximation 
ideal two dimensional graphene has a conical energy dispersion
at low energies and the Fermi surface ($E=0$) is formed by 
six points.  This low energy region can be described in terms of
a ${\vec k}\cdot {\vec p}$ theory whose mathematical structure  is very
similar to the Dirac theory for massless fermions\cite{Semenoff84}, which
 yields   physical insight. 
The ${\vec k}\cdot {\vec p}$  has also been worked out
for edge states in ideal ribbons\cite{Brey06}.
The TB approach provides   natural energy and momentum cutoffs to
the  ${\vec k}\cdot {\vec p}$ theory, permits to model perturbations
at the atomic scale  and is a good preliminary step
towards ab-initio calculations \cite{Lee04}.

The rest of the paper is organized as follows. In section II we review the
transport formalism. In section III we review the
electronic structure of ideal ribbons and we study the effect of vacancies,
located at the edge of the ribbons, on their transport properties. We find that
the conductivity of undoped graphene ribbons is weakly affected by 
the presence of this kind of disorder.
In section IV and V we study transport properties of 
 square- and wedge-shaped constrictions. Whereas the former present finite
 conductance, the latter have a vanishing transmission at low energies,
 coexisting with a finite density of localized states. These states
at energies close to zero form what can be called a quantum dot in graphene. 
In section VI we discuss the validity of our approximations
and we summarize and discuss the main results of this manuscript.

\section{Transport theory}
%\label{transport}
The natural framework for transport calculations in nanoscopic devices is the
Landauer formalism. The description of electron transport within the Landauer
formalism  is based on the  assumption that transport across the highest
resistance  region is coherent, i.e., inelastic scattering is negligible there.
A more complete account can be found, e.g., in the book by Datta
\cite{Datta:book:95}. In what follows we will assume that the inelastic  mean
free path for graphene electrons is much longer than the typical dimensions of
the nanoconstrictions considered and than the ribbon widths. According to
recent theory work, inelastic scattering due to phonons is very inefficient in
graphene ribbons so that  our assumption seems to be met even at room
temperature \cite{Gunlycke06}. This  makes  elastic scattering the major
contributor to resistance. In contrast to metallic systems where  the
resistance of the electrodes is negligible compared to that of the 
nanoconstriction, in the case of graphene nanoconstrictions, the low-bias
conductance of the ideal electrode is also very small, $\frac{2 e^2}{h}$. The
consequences of this are explored in what follows.

%\subsection{Coherent transport through a constriction}

\begin{figure}
[t]
\includegraphics[width=2.0in]{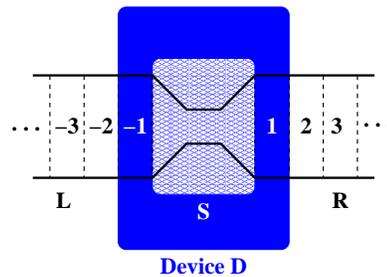}
\caption{
\label{fig:3:LDR}
 Division of system into leads L and R, device D, and scattering region S (LEFT).
 And division of operator matrices into corresponding submatrices (RIGHT).}
\end{figure}

We consider the effect of constrictions on the transport properties of an
otherwise ideal ribbon (Fig. \ref{fig:3:LDR}).  The system has three regions:
the central region (or device) where the constriction is  located and the left
and right leads. The latter are  described as semi-infinite  one-dimensional
perfect ribbons of finite width, characterized by  the number of atoms in the
unit cell, $N$. We only consider ribbons with  zig-zag edges as the ones
studied in previous works \cite{Fujita96,Wakabayashi99,Sigrist2000}.  We
consider constrictions of various shapes, from the removal of a single atom or
a few atoms on the edge to square-shaped and wedge-shaped constrictions.

The introduction of the constriction breaks the translational symmetry of the
perfect lead so that, in general, electrons incident on a given band will be
either reflected or transmitted into other bands after hitting the
constriction. The square of the transmission amplitude  $t_{nm}$
 gives the probability of an
incoming mode  $m$ to be scattered on an outcoming mode $n$.
The
Landauer formula links  the overall transmission 
  \[T(E) = \sum_{m,n}\|t_{nm}\|^2 = {\rm Tr}[t^\dagger t]\] 
with the linear conductance  
  \[G(E_F) = \frac{2e^2}{h} T(E_F)\]
where $E_F$ is the Fermi energy.
%and the current at finite voltage is given by 
%  \[I = \frac{2e^2}{h} \int {\rm d}E \,  T(E).\] 

For completeness, we review the basic steps in the calculation of the
transmission $T(E)$ using single-particle Green's functions as routinely done 
in nanoelectronics \cite{Datta:book:95}.  
In the TB approach used here, the Hamiltonian matrix is straightforwardly
obtained  for a given atomic structure. The leads are characterized by  a
unit cell with $N$ atoms and a propagation direction which we take along the
$x$ axis. When written in blocks $ N \times N$,
the Hamiltonian of the leads is
a semi-infinite tridiagonal matrix, with intra-cell blocks  $H_0$ and an
inter-cell first-neighbor coupling $V$. 
Our starting point is the partition
of the Hamiltonian of the infinite system in 3 regions.
The choice of the boundaries between device and leads is done so that
left and right electrodes are not directly coupled. 
For convenience the device region is
always chosen so that the left and right boundaries are given by one of these
units cells.  Therefore, the coupling between the surface unit cell of the
left (right) electrode to the right (left) boundary of the device is given by
the same inter-cell matrix $V$.   
According to this scheme, the Hamiltonian matrix set is divided into
submatrices as follows:
\begin{eqnarray}
  \label{eq:3:H_LDR}
  H &=& \left(
    \begin{array}{lll}
      H_{\rm L} & H_{\rm LD} & 0 \\
      H_{\rm DL} & H_{\rm D} & H_{\rm DR}\\
      0 & H_{\rm RD} & H_{\rm R}
    \end{array}
  \right),
\end{eqnarray}
Here $H_{\rm D}$ is a finite size square matrix with range 
equal to the number of atoms
in the device, $N_d$. 
In contrast, $H_{\rm L,R}$ are infinite size square  matrices
describing
the semi-infinite electrodes and $H_{\rm DL}$ and $H_{\rm LD }$ are infinite 
rectangular matrices. 
The Green function operator,  defined as 
\begin{equation}
% \label{eq:3:DefG}
( E - H ) G(E) = 1
\end{equation}
can also be divided into submatrices as:
 \begin{equation}
  \label{eq:3:G_LDR}
  G = \left(
    \begin{array}{lll}
      G_{\rm L}    & G_{\rm LD} & G_{\rm LR} \\
      G_{\rm DL} & G_{\rm D}    & G_{\rm DR} \\
      G_{\rm RL} & G_{\rm RD} & G_{R}
    \end{array}
  \right).
\end{equation}

After simple steps, it is possible to 
write the Green function of the device as:
\begin{equation}
  \label{eq:3:G_D}
  G_{\rm D}(E) = ( E - H_{\rm D} - \Sigma_{\rm L}(E) - \Sigma_{\rm R}(E) )^{-1}.
\end{equation}
where $\Sigma_{\rm L,R}$ are the so called self-energy $N_d\times N_d$ matrices
given by
\begin{eqnarray}  
  \label{eq:3:Sigma_L,R}
  \Sigma_{\rm L}(E) := H_{\rm DL} \, g_{\rm L}(E) \, H_{\rm LD} 
  \nonumber \\
  \Sigma_{\rm R}(E) := H_{\rm DR} \, g_{\rm R}(E) \, H_{\rm RD}
\end{eqnarray}
The  selfenergies $\Sigma_{\rm L}(E)$ and $\Sigma_{\rm R}(E)$
describe the effect of the electrodes on the electronic structure of the {\em
device}. The real part of the self-energy  
results in a shift of the device levels whereas the imaginary part provides a
lifetime. 
The device self-energies are given by the Green's functions of the semi-infinite isolated leads
$g_{\rm L}(E) = ( E - H_{\rm L} )^{-1}$ and $g_{\rm R}(E) = ( E - H_{\rm R}
)^{-1}$ projected into the device region  by the coupling of the leads to the
device $H_{\rm DL}$ and $H_{\rm RD}$.
In contrast to the Green function of an infinite system with translational
invariance,  
the calculation of the Green function of a
semi-infinite system with a surface is non-trivial.

We can write  the surface part of the semi-infinite Green function 
$g_{\rm L,R}(E)$ of the electrode as:
\begin{equation}
  \label{g_22}
  g_{\rm L,R}(E)|_{\rm surface} = \frac{1}{E - H_0 - \Sigma_{\rm l,r}(E)},
\end{equation}
where $\Sigma_{\rm l,r}$ 
is a self-energy (different from $\Sigma_{L,R}$)
 that accounts for the effect of the rest of the semi-infinite
chain on the first unit cell. 
In one dimension it is possible to derive a
recursive relation that yields a self-consistent Dyson equation for
this self-energy:
\begin{equation}
  \label{Sigma}
  \Sigma_{\rm l,r}(E) = V \frac{1}{E - H_0 - \Sigma_{\rm l,r}(E)} V^\dagger.
\end{equation}

The coupling matrices $\Gamma_{\rm L}(E)$ and $\Gamma_{\rm R}(E)$ are defined
as the difference between the retarded and advanced selfenergy of the leads
projected into the device by the coupling $H_{\rm DL}$ and $H_{\rm RD}$:
\begin{equation}
  \label{eq:3:Gamma_L,R}
  \Gamma(E)_{\rm L(R)} = i \left( \Sigma_{\rm L(R)}(E) - \Sigma_{\rm L(R)}^\dagger(E) \right).
\end{equation}

With all these ingredients, we can compute the transmission
using the result\cite{Datta:book:95}:
\begin{equation}
  \label{eq:3:caroli}
  T(E) = {\rm Tr}[\Gamma_{\rm L}(E) G_{\rm D}^\dagger(E) \Gamma_{\rm R}(E) G_{\rm D}(E)].
\end{equation}
Therefore, for a given system, we first compute the electrode surface Green
function (\ref{g_22}) by solving the Dyson equation (\ref{Sigma}). This permits
to compute the device self-energy (\ref{eq:3:Sigma_L,R}), the device Green
function (\ref{eq:3:G_D}) and the coupling matrices (\ref{eq:3:Gamma_L,R}).
 The final step is the calculation of the
transmission function. 

\section{Ideal and weakly defective zig-zag ribbons}

\subsection{Ideal ribbons}
\begin{figure}
\includegraphics[width=3.in]{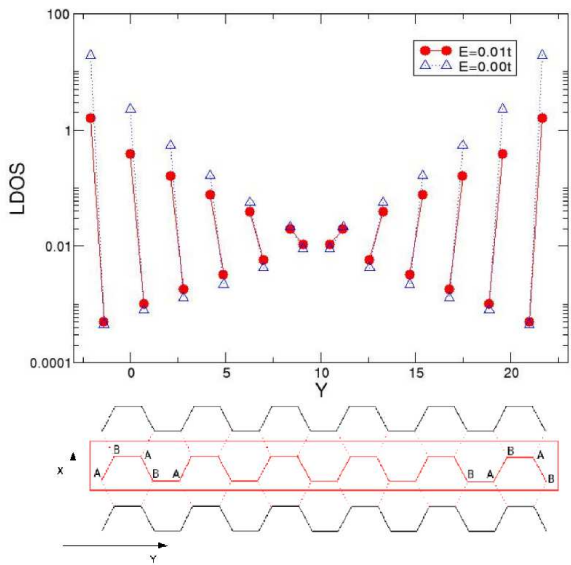}
\caption{
\label{maps-bands}
Upper panel: Local density of states of the ideal 24 atoms-wide ribbons  
for $E=0$ (circles) and $E=0.01t$ (triangles). Notice the vertical logarithmic
scale.  Lower panel: atomic structure of a a section of a $N=24$ zig-zag ribbon.}
\end{figure}

In this section we briefly review the electronic structure and transport
properties of both ideal and weakly defective  zig-zag-edged
ribbons.  We consider ribbons with two different widths
whose unit cells are composed of $N=24$ and $N=48$ atoms, respectively. 
The super-cell unit of the $N=24$ ribbon is shown in Fig.
 (\ref{maps-bands}b).
The honeycomb 
lattice of 2D infinite graphene 
can be generated by a triangular lattice of unit cells with two atoms, 
labeled $A$ and $B$. Therefore, the honeycomb lattice
is formed by two interpenetrating sub-lattices, $A$ and $B$. The
first neighbors of atoms in the lattice $A$ belong to sub-lattice $B$ and
vice versa. This underlying structure is responsible for most of the peculiar
features  of graphene electronic structure. Nano-ribbons inherit these
properties and, in the case of zig-zag nano-ribbons, 
the top and bottom edges 
belong to atoms on different sublattices \cite{Brey06}. 

\begin{figure}
[t]
\includegraphics[width=2.7in]{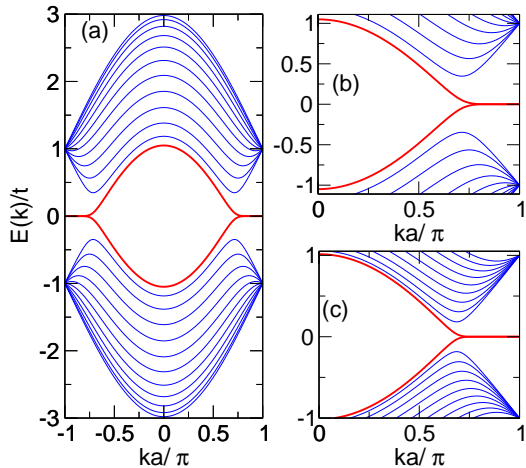}
\caption{
\label{fig-band}
(a) Bands of the ideal 
 zig-zag ribbon with 24 atom unit cell. (b) Detail of the low energy sector.
 (c) Detail of the low energy sector for the 48 atom unit cell  }
\end{figure}

Figure (\ref{fig-band}) shows the band structure for  the
$N=24$ case [Figs. \ref{fig-band}(a) and  \ref{fig-band}(b)] and for
the $N=48$ case [only the low energy region is now shown in
Fig. \ref{fig-band}(c)].  The energy units  
are given in terms of the hopping parameter $|t|\simeq 3$eV
which is the only energy scale in the Hamiltonian. 
There is a perfect electron-hole symmetry which sets the Fermi energy 
to zero for  half filling.  Notice that the density  of bands per energy
interval is largest for the wider ribbon, as expected. 
We can distinguish three different regions, (i) $|E|<\Delta_1$, (ii)
$\Delta_1<|E|<|t|$
and (iii) $|E|>t$, where $\Delta_1$ is the minimum energy of the 
second sub-band closest to $|E|=0$.
This energy scale $\Delta_1$ is associated to the finite width 
of the ribbon and it  decreases as $N^{-1}$. For the cases considered here,
$N=24$ and $N=48$, $\Delta_1=0.35$ t and $\Delta_1=0.18$t respectively.
In the $N=\infty$ limit $\Delta_1$ goes to zero, as expected for 
two-dimensional graphene.  

The density of extra carriers that needs to be injected in the ribbon so that the
Fermi energy hits the second subband, $E_F=\Delta_1$, exceeds the upper
experimental limit reached so far \cite{Bunch05}, $\delta n=10^{13}$ cm$^{-2}$ .
Electrical doping  up to the second subband could be possible for wider ribbons.
 Although only the lowest energy region $|E|<\Delta_1$ could be accessible
experimentally for the narrow ribbons considered here,  some insight can gained
also by analyzing the effect of elastic scattering on the transmission at energies
in the other two regions. The  bands immediately above $\Delta_1$ present two
positive (and two negative)  momenta at a given energy. This results in the
doubling of the number of conducting channels at a given energy, as long as
$|E|<t$. Higher in energy,  $|E|>t$, we find simple parabolic bands which yield
one channel per band as in the case of III-V semiconductors.

% We can estimate the density of extra carriers
%injected in a 2D graphene sheet whose chemical potential is shifted up to
%$\Delta_1$, for different values of $N$. 
%From the linear density of states of ideal 2D graphene, 
%$\rho(E)=\frac{|E|}{2 \pi t^a a^2}$ we obtain, upon trivial integration,
% \begin{equation}
%\delta n = \frac{\Delta_1^2}{4 \pi t^a a^2}
%\end{equation}.
%(there might be a LINEAR FACTOR in all this !!!). 

The energy bands  closest to $E_F$ are flat for $|k|$ larger than a critical
wave-vector $k_c$, and
present dispersion otherwise. The cutoff $k_c$ 
is a decreasing function 
of $N$ which tends to  $k_c a/\pi=2/3$ in the  
 $N=\infty$ limit\cite{Fujita96}. 
In  figure  \ref{maps-bands} we plot the local density of states across a super-cell unit
for the ribbon with $N=24$ and for two different energies, $E=0$ and $E=0.01$t. 
Atoms $A$ and $B$ in the same unit cell are joined by a straight line. 
In both cases the LDOS is peaked on the edges, although the contrast is larger
for the $E=0$ case, which corresponds to the 
dispersionless states.  The LDOS presents a peculiar oscillation, related to the
sub-lattice structure. Starting from the left edge,
the LDOS peaks only in $A$, the weight on the first $B$ atom being very small.
As we move towards the center the weight on $A$ atoms 
decays exponentially whereas the weight on the $B$ atoms increases
exponentially. In  the center the weight on the $A$ and $B$ atoms is comparable
and the weight on the $B$ atoms becomes dominant as we move towards the
opposite edge, in very good agreement with the $\vec{k}\cdot\vec{p}$ theory
\cite{Brey06}. As a result, the density of states, disregarding the
sub-lattice index, is peaked at the edges, which permits to refer to the lowest
subband states as "edge states".

It is apparent from Fig. \ref{maps-bands} that 
the $A$ and $B$ edge states become coupled in the middle of the ribbon. 
Importantly, the structure
of the current operator is the same as 
the hopping part of the Hamiltonian, coupling atoms
of the different sub-lattices. Therefore, the current density of a given state,
evaluated in a unit cell with two atoms, 
is proportional to the
product of its  $A$ and $B$ components, $\psi_A(y)$ and $\psi_B(y)$. 
In the continuum limit the current associated to the wave function 
$\Psi^{\dagger}=\left(\psi_A^*,\psi_B^*\right)$  reads
$j_x(y)\propto \left(\Psi^{\dagger} \sigma_x \Psi\right)$,
 where
$\sigma_x$ is the Pauli matrix acting on the $AB$ 
space\cite{Katsnelson06,Beenakker06,Cheianov06}.
From  figure (\ref{maps-bands})
we expect that the product $\psi_A(y)\times\psi_B(y)$
to take similar values in the edges\cite{Brey06} and 
in the center of the
ribbon. The resulting picture is the following: whereas the
charge density of low energy states,
proportional to $\rho(y)=|\psi_A|^2+|\psi_B|^2 $,
 is peaked on the edges, their  current density $j_x(y)$ is more
 homogeneously distributed accross 
the  ribbon. This picture is substantially 
different from  that of non-relativistic electrons  with
scalar wave functions $\phi(x,y)=\psi(y) e^{ik x}$ for which the charge density  
$\rho(y)=|\psi(y)|^2$ and the current density  
$j_x(y)\propto \phi \partial_x\phi^*-\phi^* \partial_x\phi\propto k \rho(y)$ have the
same profile. 
The consequences of this  difference between the current and charge densities
will become apparent later and are one of the results of this work.

\subsection{Defective edges}

\begin{figure}
[t]
\includegraphics[width=2.7in]{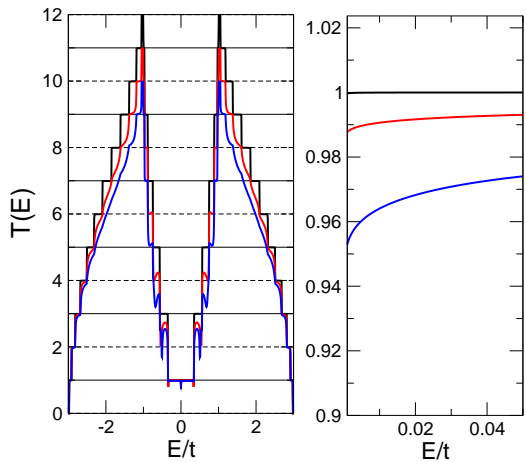}
\caption{
(Color on-line).
\label{Vacancy}
(a) Transmission spectrum for ribbons with $N=24$ atoms in the ideal case (black),
removing 1 atom from the edge (red), and removing 2 atoms on the top and bottom edges (blue).
(b)  Detail
of the transmission in the low energy sector. The effect of removing 1 or 2
atoms on the border is very small in the lowest conductance plateau. }
\end{figure}

We now study the effect of a single vacancy in the edge(s) of the ribbon on 
the transport properties.  From the formal point of view this is done using the
approach described in the previous section. The sector of the ribbon where the
vacancy is located is treated as the device. 
In  Fig. \ref{Vacancy} we plot $T(E)$ for three cases:  
Ideal case,  one atom missing and two atoms missing. 
As everywhere else in the text, this is the transmission per spin channel. 
Since we assume that time-reversal symmetry is not broken,  the total
transmission should be multiplied by a factor of 2 to account for the spin
degree of freedom.
All of them display electron-hole
symmetry around $E=0$.  As in the case of the energy bands,
the $T(E)$ curves have three different regions. 
In the large $|E|$ region, the transmission for the ideal ribbon
 $T_{\rm ideal}(E)$ is quantized according to
the usual law for non-relativistic electrons and holes:
$n e^2/h$, where $n$ is the number of bands that intersect the Fermi energy
for $k>0$. In this case $n$ goes from 1 to 12 in steps of 1, 
consistent with the parabolic dispersion away from the Dirac cones.
This large $|E|$ region is not likely to 
be reached experimentally since it would imply
a huge depletion of the charge, but is interesting from the conceptual point of
view.
In the intermediate region ($t<|E|<\Delta_1$)  $T_{\rm ideal}(E)$ changes 
according to the rule $(2n+1) e^2/h$.
This result has been previously obtained by Peres et al.
(\onlinecite{PRBGuinea06}).  The factor of two comes from the shape of the
bands in this region, as discussed above. 
The transmission is maximal at the energy where  the
"Dirac" ladder and the non-relativistic ladder
meet at $E\simeq t$. In the relevant low energy sector, where
we will focus our attention from now on, the transmission of the
ideal ribbon is one.

The effect of scattering introduced in the ribbons
affects very differently the transmission in the
different energy regions. The  removal of a {\em single}
atom on the edge erases the
$2n+1$ ladder in the intermediate energy region for the $N=24$ ribbon 
[see Fig. \ref{Vacancy}(a)]. 
This points towards
a very difficult experimental verification of the $2n+1$ transmission
ladder in these narrow ribbons. 
In contrast, in the case of $N=48$ the conductance
remains quantized in the $2n+1=3$ plateau after the removal of 1 atom in each
edge, but the higher energy plateaus  disappear.

In contrast with the higher energy subbands ($|E|>\Delta_1$),
 the plateau in the
low energy sector is only weakly affected by the removal of atoms 
in the edges, even for the narrower
ribbon ($N=24$) as can be seen
in Fig. \ref{Vacancy}(b). 
The effect on
the transmission in the experimentally relevant
low energy sector is below two  percent for the removal of one  atom
on one edge and below five percent when one atom is  missing on both 
edges. In contrast, the removal of a single atom in the central part of the
ribbon (not shown) has a much larger influence on the low energy transmission.
These results are compatible with the fact that the current density carried by
edge states is spread along 
the central region. The robustness of the transmission in the low energy sector and its weakness in
the intermediate energy sector are relevant results since atomic size
fluctuations in the edges are unavoidable in real samples.

\section{Square Constrictions}

\begin{figure}
[hbt]
\includegraphics[width=2.in]{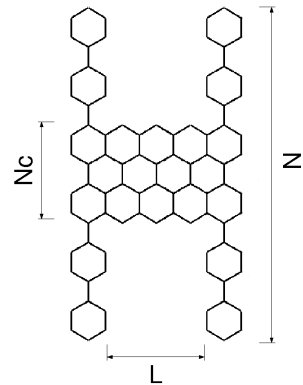}
\caption{\label{structure-square1}
(Color On-line). Structure of the square-shape nanoconstriction with 3
characteristic lengths, $N$, $N_c$ and $L$. }
\end{figure}

We now present results for square-shaped nanoconstrictions as those shown in
Fig.  \ref{structure-square1}. We choose this shape because it permits a
comparison with a simple model for electrons in parabolic bands.  It also
seems possible to obtain analytical expressions for the transmission curve 
for square constrictions  using analytical approach 
along the lines of previous  work \cite{Katsnelson06,Beenakker06,Brey06}. 
Another feature
of square constrictions is the presence of armchair edges joining the zig-zag
edges of the wide and narrow regions.
We consider square constrictions with top-bottom symmetry,  like those in the
Fig. (\ref{structure-square1}), characterized 
 by three  lengths: The width of
the electrode and the width of the constriction, $N$ and $N_c$, both measured in
units of the number of atoms of the super-cell unit,  and the length of the
constriction $L$, measured in units of $a$, the graphene lattice parameter.

\begin{figure}
[hbt]
\includegraphics[width=2.8in]{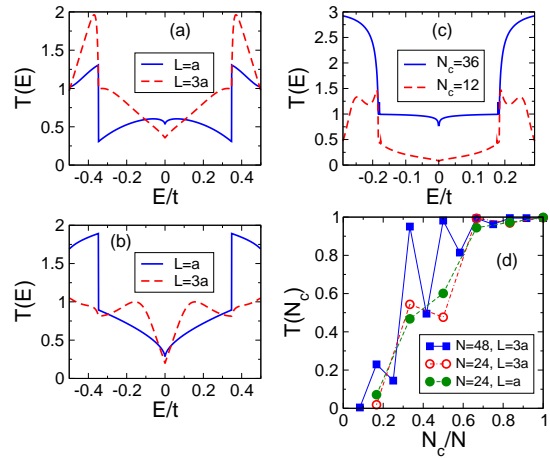}
\caption{\label{Trans-sq-2}
(Color On-line)
Square constrictions. 
(a) $T(E)$ curves for the $N=24$, $N_c=12$ and
two values of $L$. (b) $T(E)$ curves for the $N=24$, $N_c=8$ and
two values of $L$. (c) $T(E)$ curves for $N=48$, $L=3a$ for 
two constrictions with different $N_c$. (d) $T(E=0.05t)$ as a function of
the aspect ratio of the constriction $N_c/N$ for 3 families of constrictions. 
}
\end{figure}

In Fig.  (\ref{Trans-sq-2}) we show transmission curves for a variety of square
constrictions. In Figs. \ref{Trans-sq-2}(a) and (b) we show results for electrodes
with $N=24$ atoms, whereas in \ref{Trans-sq-2}(c) we show results for wider
electrodes with $N=48$. The two curves in Fig. \ref{Trans-sq-2}(a)
permit to compare
the transmission for two structures   
with the same aspect ratio $N_c/N=12/24$ but different nanoconstriction length 
$L$. 
Analogously, Fig. \ref{Trans-sq-2}(b)
shows two $T(E)$ curves with the same aspect ratio 
$N_c/N=8/24$ and different $L$. 
In Fig. 
\ref{Trans-sq-2}(c) we show two $T(E)$ curves for a wider electrode, 
keeping the length of the constriction fixed to $L=3a$  and changing the aspect
ratio.  
The transmission corresponding to the lowest subband $|E|<\Delta_1$, 
is significantly reduced with respect to the ideal ribbon $T=1$, 
but remains finite ($T(0)\simeq 0.18$) even for the
narrowest and longest constriction. 

In order to highlight the peculiar properties of graphene, 
we give two arguments, which turn out to be inapplicable, to expect a
vanishing transmission at zero energy.
 First, 
since low energy transmission is associated to edge
states and given that the edges of the electrode and the constriction are not
connected, the transmission would vanish as $N_c$ becomes much smaller than $N$. 
This  argument fails because the current density of edge states has a sizable
contibution in the center of the ribbon, which is smoothly connected to the
constrictions considered in Fig.  (\ref{Trans-sq-2}).
The second argument is based on the behavior of a square constriction with 
parabolic-band electrons. In that case, the
different width of the semi-infinite ribbon
$W_r$ and the constriction  $W_c$  yields different band minima
so that an energy gap is created in the narrow region. This can be
  modeled with square-shaped barrier  of height 
$V_0 =\frac{\hbar^2}{2 m*}\left(\frac{1}{W_c^2}-\frac{1}{W_r^2}\right)$ 
and   length $L$,
$m^*$ being the effective mass of the electron. 
The zero energy transmission of a
square barrier is always exponentially vanishing with the length,
 in contrast to our results. 
This second argument to expect a vanishing transmission at zero energy
also fails because the situation for graphene ribbons is
 very different: Both wide and
narrow sectors of the system have zero energy edge states. 
In other words, it is not possible to create a gap in the low energy region by
changing the width of the system so that it is not possible to create a barrier
for the carriers in the lowest energy sub-band by geometrical means. 
Interestingly, the transmission through a square potential has been calculated,
in the case of infinitely wide ribbon, and the transmission does not vanish for
electrons incident perpendicular to the barrier \cite{Katsnelson}.

The results in Fig. \ref{Trans-sq-2}(a) and (b) suggest that the low energy
transmission is an increasing function of the constriction width
$N_c$ and and fairly insensitive to constriction length $L$. 
In Fig. \ref{Trans-sq-2}(d) we plot the transmission at 
a fixed energy $E= 0.05$t as a function
of $N_c/N$ for three families of constrictions. Square symbols correspond to a wide
ribbon with $N=48$, and $L=3a$, whereas  solid (open) circle symbols correspond to a narrow
ribbon with $N=24$ and $L=a$ ($L=3a$). 
It is apparent that as the aspect ratio increases, 
the transmission increases in average, with super-imposed  oscillations. 
This behavior is
different from the analogous curve for transmission as a function of the
barrier height $T(E,V_0)$ for parabolic-band electrons in the tunneling regime
($E<V_0$), where the curve does not present oscillations. 
In the $E>V_0$ regime
the $T(E,V_0)$ curve can present oscillations \cite{Datta:book:95}.
Therefore, the $T(N_c/N)$ curve looks
closer to the latter case, even for very small energy. 
This is related to the lack of a gap in the density of states. 
It is also interesting that  for both wide and narrow electrodes the
transmission saturates to 1 for the same value of $N_c/N\simeq 0.7$. 
As a final remark, only in the case of atomically narrow constrictions the
transmission becomes almost zero. In summary, square constrictions are not very
efficient in blocking electronic transport in zig-zag graphene ribbons.

\section{Wedge-shaped constrictions}

\begin{figure}
[hbt]
%\begin{graphicsarray}
\includegraphics[width=0.98in]{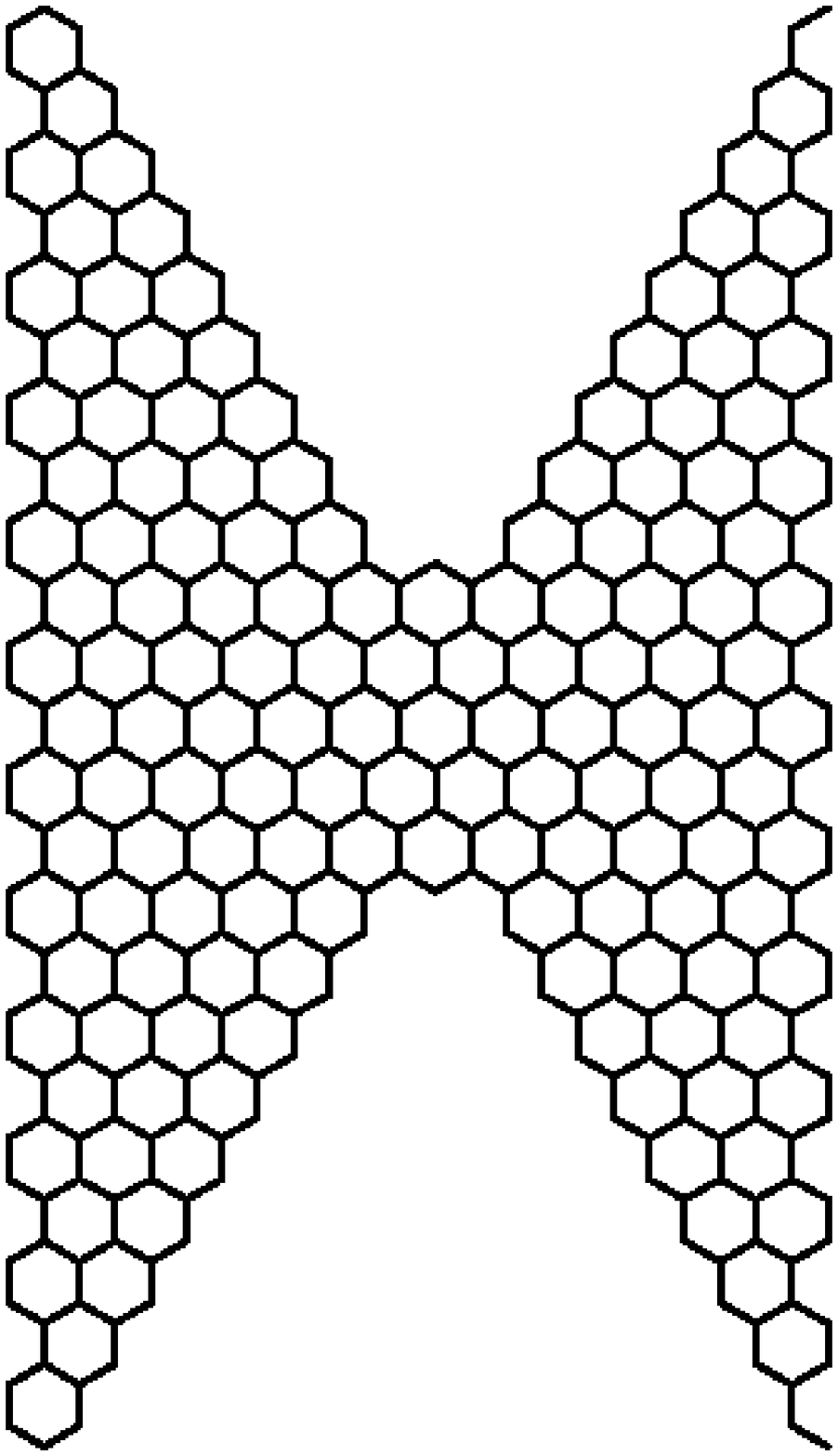} \,\,\,
\includegraphics[width=0.98in]{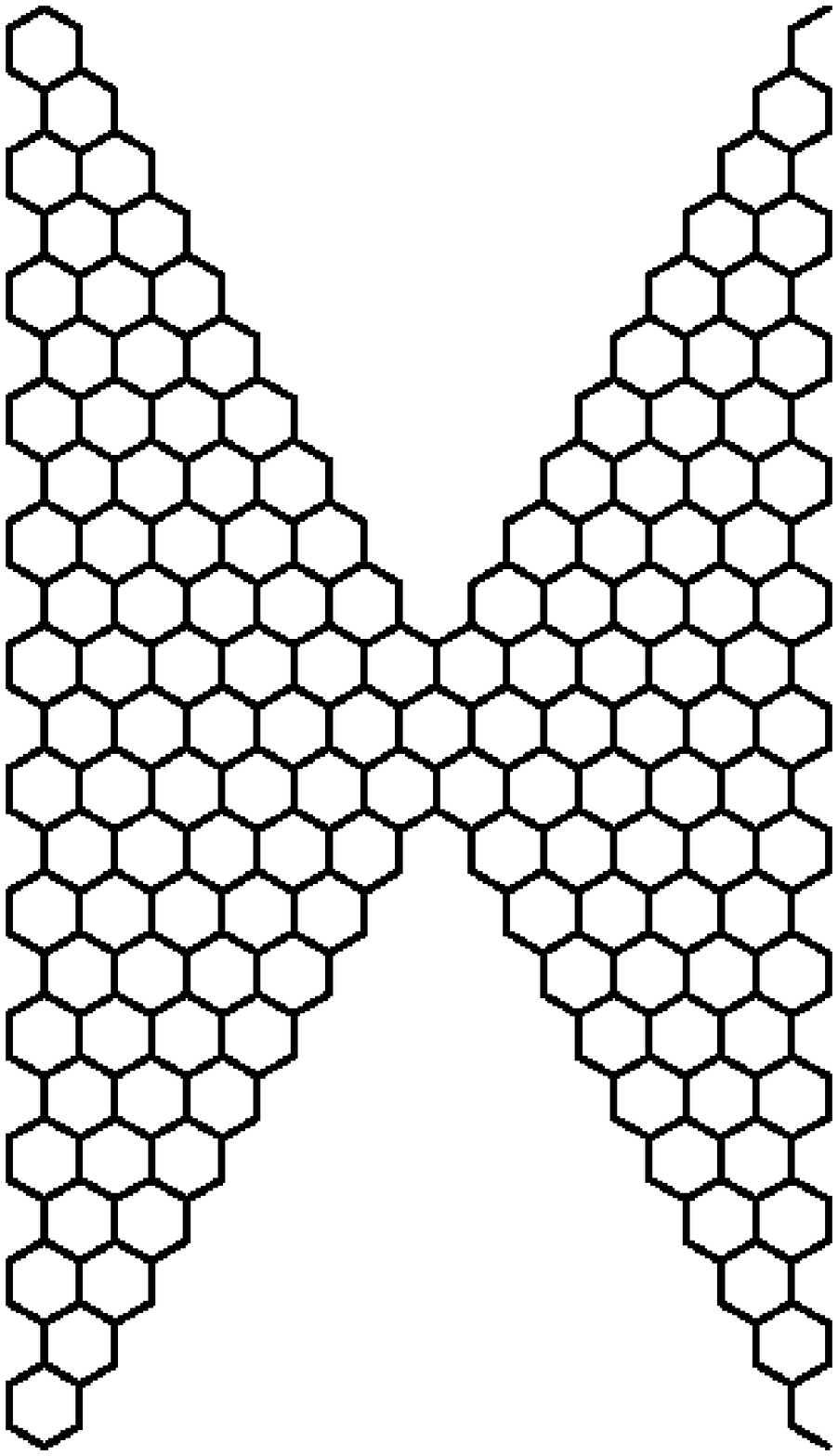}\,\,\, 
%\end{graphicsarray}
\caption{
\label{structure-wedge}
Schematic atomic structure of wedge-shaped constrictions}
\end{figure}

We now turn our attention to a wedge-shaped constriction,
shown in Fig. \ref{structure-wedge}.
As opposed to the square constrictions considered above, 
wedge-shaped constrictions are always delimited by  zig-zag edges.
It can also be expected that wedge-shaped constrictions are
more stable than square constrictions, although further work should clarify
this point. 
The structures considered have a top-bottom symmetry, 
in contrast to the asymmetric structures  considered in earlier
work \cite{Sigrist2000,Wakabayashi2001}. 
Naively,  the wedge-shaped constriction would allow for adiabatic  transport
since the zig-zag edge gets never interrupted. In contrast, we see
that for ribbons with $N=48$ atoms the constrictions
shown in Fig.  \ref{structure-wedge}
yield a vanishing transmission  in the whole low energy sector
$|E|<\Delta_1$.
We have verified that these results are robust against the removal of
a single atom from one of the edges. 
Surprisingly,  these smooth nano-constrictions 
seem to be more disruptive for low energy transmission than
the  abrupt nano-constrictions considered in the previous section. 

\begin{figure}
[t]
\includegraphics[width=2.7in]{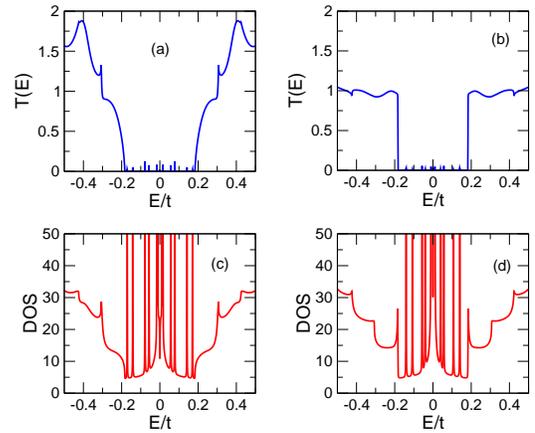}
\caption{
\label{Trans-wedge}
(Color On-line). (a) and (b) $T(E)$ for
the structures shown in Fig. \ref{structure-wedge} (blue). (c) and (d) 
Corresponding density of states projected on the whole constriction region (red).}
\end{figure}

\begin{figure}
[b]
\includegraphics[width=2.5in]{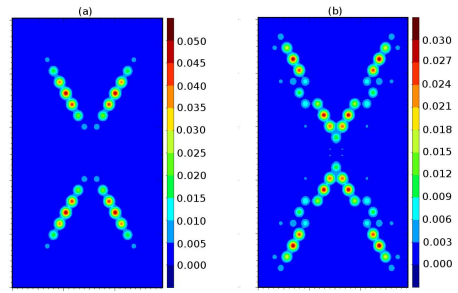}
\caption{
\label{LDOS-wedge}
Local density of states for the right wedge structure
shown in Fig. \ref{structure-wedge} at energies $E=0$ (a) and $E=0.04t$ (b).}
\end{figure}

Interestingly, the vanishing transmission is {\em not} associated to a
vanishing density of states in the device region, as shown  in Figs.
\ref{Trans-wedge}(c) and (d).  The peaks in the
DOS in the energy region with zero
transmission is associated with low energy states very weakly coupled, if at all,
to the electrodes. The width of the peaks in the DOS is limited by the
numerical broadening used in the calculation. 
In Fig. \ref{LDOS-wedge} we show
a contour map of the DOS at $E=0$  and $E=0.04t$, the energy at which the
first peak of the device DOS is located for the structure shown on the right 
 in Fig. \ref{structure-wedge}. 
 The maps reveal localization on the four edges of the constrictions,
 reflecting the fourfold symmetry of the device. 
Notice that the LDOS vanishes at the boundary with the
electrodes, which explains the vanishing transmission. This LDOS
could be observed in STM experiments 
\cite{Niimi05,Kobayashi05}.
% if fabrication of these structures were possible. 

Whereas the probability density across one of the four edges
has a bell shape for $E=0$, it has a node 
for the finite energy case. 
The properties of these low energy
non-conducting states in this wedge-shaped structures are very similar to those of
zero-dimensional confined states. 
Confinement in semiconductor heterostructures  is associated with the existence
of an energy gap. The absence of such a gap in graphene makes it necessary to
look for different strategies to confine electrons 
\cite{Katsnelson,Efetov,Peeters}.  In this regard, 
nano-constrictions like the ones studied in this section
can behave like quantum dots and might permit the
study of Coulomb Blockade and Kondo physics in graphene structures.  
The physical origin of these bound symmetric wedge states (SWS) might be
related to the formation of Kekul\'e vortex structures discussed in the case of
asymmetric wedge states \cite{Sigrist2000,Wakabayashi2001}. In the asymmetric
case, the suppression of the transmission occurs for very narrow energy windows,
since one of the edges is not perturbed. 
It would be desirable to study whether the vanishing tranmission that we have
found is related to the vanishing tranmission obtained
analytically\cite{Katsnelson} for the square barrier potential, in infinitely
wide ribbons, for incidence angles different from zero. 
In summary, symmetric wedge
constrictions result in a gap in the transmission curve for $|E|<\Delta_1$,
 yet with a finite density of states in that interval, featuring very narrow
 peaks that mimic a discrete spectrum of confined states.

\section{Discussion and Conclusions}

Here we discuss the validity of the approximations  assumed for our  calculations. 
Real samples could present several features absent in the idealized
ribbon considered here. On one side, the presence of chemical impurities, like  hydrogen both
in the edge and in bulk, water molecules, oxygen, etc., will affect the electronic
structure and transport properties of the system. In a sufficiently clean and
ultra-high vacuum environment the effect of chemical impurities could be
negligible.  In experiments, the graphene ribbon is deposited
on an insulating
substrate, which has not been included in our calculation. First-principles
calculations indicate that the effect of
the substrate is weak \cite{APL}. 

The use of a single-particle model certainly fails, in a trivial sense,
if the there are  deviations from charge neutrality which would
make it necessary to perform a self-consistent calculation including, at
least, the Hartree contribution. This is certainly achieved by density functional
calculations. 
We have verified that, for the structures considered here, 
the electronic density does not deviate significantly from
charge neutrality.
The single particle approach also
fails if the electron liquid happens to be different from a Fermi liquid, in which
the low energy quasiparticles have the same  quantum numbers than the free
electrons. Such a scenario has been considered for two dimensional graphene
\cite{Geli96}. A priori, this is a serious issue in a one dimensional system, where
the Fermi liquid state is not stable with respect to 
electron-electron interaction  and a Luttinger liquid is expected. The same
statement applies for nanotubes. However, 
%the experimental confirmation of
%Luttinger liquid behaviour in nanotubes remains elusive so far and single
the single-particle Fermi liquid picture describes most of the experimental results in
nanotubes and the same can be expected for graphene ribbons. 

We have used the same on-site energy ($0$) and hopping $t$
for the edge and the inner atoms.
This is an approximation, but the results do not change significantly in the
case of ideal ribbons \cite{Ezawa06}.
Another limitation of our model is the one orbital approximation. Ab-initio
calculations, including all the atomic orbitals,  show that the dangling bonds 
present on the edge atoms form a flat  band close to the Fermi energy 
\cite{Lee04}.  We have also ignored the spin degree of freedom. The flat band
at zero energy is expected to spin split due to spontaneous magnetization
induced by the Coulomb repulsion \cite{Lee04,Guineamagnetism}. This interesting issue
deserves more theoretical and experimental work. We have also ignored
spin-orbit interaction, which is very small in carbon, although it has
attracted some interest \cite{Spin-orbit-NT,Spin-orbit-dani}, 
 in part due to the Spin Hall effect \cite{Spin-orbit}.

In summary, 
we have studied coherent transport in graphene nano-ribbons with zig-zag edges.
The electronic
structure of the ribbons is described with a simple one orbital tight-binding
approximation. We have focused on narrow ribbons ($W< 5 nm$ )
for which only the lowest energy sub-band is expected to be experimentally relevant. 
 Our results could be summarized as follows:
{\em i)} The low-energy transport properties are robust with respect
to isolated vacancies on the edges.  
{\em ii)} These are also robust with respect to square-shaped
nano-constrictions. Linear conductance survives in most cases except for very
narrow constrictions.  This is at odds with the behavior of parabolic-band
electrons in similar constrictions and reflects two non-trivial features of edge
states: Their minimum energy is independent of the ribbon width and their
current density profile spreads beyond the edges of the ribbon, in 
contrast with  their density profile.
{\em iii)} In contrast to square constrictions,
 wedge-shaped nano-constrictions
result in a gap in the transmission and result in a zero linear conductance
which is related to the appearance of localized low energy edge states. These
edge states have properties similar to those of confined states in zero
dimensions. Therefore, graphene wedge shape constrictions might have properties
analogous to those of semiconductor quantum dots.
Extensions of this work will address the spin degree of freedom, topological
defects \cite{Cortijo06} and armchair edges.

We acknowledge useful discussions with M. Vozmediano,   A. Cortijo  and B. 
Valenzuela.    This work
has been financially supported by MEC-Spain (Grants FIS200402356, MAT2005-07369-C03-01
 and the Ramon y Cajal Program) and by
Generalitat Valenciana (GV05-152).  This work has been partly
funded by FEDER funds.

{\em Note added}: After  the completion of this work we have been aware of a
related theory paper\cite{Valley-valve} with transport calculations in
wedge-shape nano-constrictions similar  to that of figure 
(\ref{structure-wedge}).

\widetext

\begin{references}

\bibitem{Bunch05} J. Scott-Bunch 
{\em et al.},
%,Yuval Yaish, Markus Brink, Kirill Bolotin, and Paul L. McEuen
Nanoletters {\bf 5} 287 (2005)

\bibitem{Geim05} 
%Geim Nature 2005
 K. S. Novoselov {\em et al.}, Nature {\bf 438}, 197 (2005).
 
\bibitem{Kim05}  
%Kim Nature 2005
 Y. Zhang {\em et al.}, Nature {\bf 438}, 201 (2005).
 
\bibitem{Science06} 
C. Berger {\em et al.},  Science {\bf312}, 1191 (2006)


\bibitem{Katsnelson06}  M. I. Katsnelson, Eur. Phys. J.  {\bf 51}, 157 (2006)
%cond-mat/0512337,
%Zitterbewegung, chirality, and minimal conductivity in graphene

\bibitem{Beenakker06}
%cond-mat/0603315 
%Quantum-limited shot noise in graphene
J. Tworzydlo, B. Trauzettel, M. Titov, A. Rycerz, C.W.J. Beenakker
Phys. Rev. Lett. {\bf 96} 246802 (2006)

\bibitem{Diffusive}
%Quantum transport of massless Dirac fermions in graphene
Kentaro Nomura, A.H. MacDonald, 
%Phys. Rev. Lett. {\bf }, (2006)
cond-mat/0606589

\bibitem{Semenoff84} G. W. Semenoff, Phys. Rev. Lett{\bf 54}, 2449 (1984)
%Paper on equivalence of TB HoneyComb and 2+1 Dirac equation

\bibitem{Fujita96} 
%Fujita PRB 1996
K. Nakada, M. Fujita, G. Dresselhaus and M. S. Dresselhaus,
Phys. Rev. B{\bf 54}, 17954 (1996)

\bibitem{Wakabayashi99} 
K. Wakabayashi,  M. Fujita, H. Ajiki, M. Sigrist,
Phys. Rev. B{\bf59}, 8271 (1999)

\bibitem{Ezawa06} M. Ezawa, Phys. Rev. B 73, 045432 (2006).

\bibitem{Brey06} L. Brey and H. Fertig, 
Phys. Rev. B{\bf 73}, 235411(2006)
%cond-mat/0603107 (unpublished).
\bibitem{PRBGuinea06} 
%transport 4(n+1/2)
N. M. R. Peres, A. H. Castro Neto, and F. Guinea
Phys. Rev. B {\bf 73}, 195411 (2006)
% erratum
N. M. R. Peres, A. H. Castro Neto, and F. Guinea
Phys. Rev. B {\bf 73}, 239902 (2006)


\bibitem{Book} {\em  Physical Properties of Carbon Nanotubes}, 
R Saito, M S Dresselhaus, G Dresselhaus, World Scientific, Singapore (1998)

\bibitem{NT-Dai}
A. Javey, J. Guo, Q. Wang, M. Lundstrom and H. Dai
Nature {\bf 424}, 654 (2003)
%Ballistic carbon nanotube field-effect transistors
 
\bibitem{Jarillo1} 
% (Electron hole symmetry)
P. Jarillo-Herrero , S.  Sapmaz, C. Dekker , L.P. Kouwenhoven LP, 
van der Zant H. 
Nature {\bf 429} 389 (2004)

\bibitem{KondoNT}
% Kondo physics in carbon nanotubes
%Authors:
 J. Nygard, D. H.  Cobden, P.  E.  Lindelof
Nature {\bf 408}, 342 (2000)







\bibitem{Jarillo2} 
%Pablo, SU4 Kondo
P. Jarillo-Herrero, J.  Kong, H. van der Zant, 
C. Dekker, L.P.  Kouwenhoven,
S.  De Franceschi Nature, {\bf 434}, 484 (2005)

%\bibitem{Jarillo3} 
%Pablo, Supercurrent transistor
%P. Jarillo-Herrero , J. van Dam , L. P.  Kouwenhoven 
%Nature {\bf 439} 953, (2006) 


%{\bf EXPERIMENT on EDGE STATES}

\bibitem{Niimi05} Y. Niimi, T. Matsui, H. Kambara, K. Tagami, M. Tsukada, and H. Fukuyama, Appl.
Surf. Sci. {\bf 241}, 43 (2005).
% STM experimental dI/dV on edges

\bibitem{Kobayashi05} Y. Kobayashi, K. I. Fukui, T. Enoki, K. Kusakabe, and Y. Kaburagi, 
 Phys. Rev. B {\bf 71}, 193406 (2005).
% STM experimental dI/dV on edges

%{\bf  THEORY PAPERS ON RIBBONS}





% kp theory for edge states

%\bibitem{GquantGaAs}
%B. J. van Wees, H. van Houten, C. W. J. Beenakker, J. G. Williamson, L. P.
%Kouwenhoven, D. van der Marel, and C. T. Foxon Phys. Rev. Lett. {\bf 60}, 848
%(1988)


%{\bf TRANSPORT PAPERS}


\bibitem{Sigrist2000} K. Wakabayashi and M. Sigrist, Phys. Rev. Lett.{\bf84},
3390 (2000)

\bibitem{Wakabayashi2001}
 K. Wakabayashi, Phys. Rev. B{\bf 64} 125428 (2001)


\bibitem{Lee04} 
% Ab initio paper
Y. Miyamoto, K. Nakada, and M. Fujita, Phys. Rev. B {\bf 59}, 9858 (1999);
 
H. Lee {\em et al.}, Chem. Phys. Lett. {\bf398}, 207 (2004)

%Graphitic ribbons without hydrogen-termination: Electronic structures and stabilities
Takazumi Kawai, Yoshiyuki Miyamoto, Osamu Sugino, and Yoshinori Koga
Physical Review B {\bf 62}, R16349 (2000)


\bibitem{Datta:book:95}  S. Datta, {\em Mesoscopic Transport }. 


\bibitem{Gunlycke06}
%Room Temperature ballistic transport in graphene nanostrips
D. Gunlycke, H. M. Lawler, C. T. White, cond-mat/0606693

%\bibitem{Dirac} G. W. Semenoff, Phys. Rev. Lett{\bf 53},  2449(1984)

%New stuff


\bibitem{Cheianov06} 
V. V. Cheianov,  V. Falko, cond-mat0603624
%Falko n-p junction



%KP papers

 

%CONFINEMENT PAPERS

\bibitem{Katsnelson} 
%Klein paradox in graphene
M. I. Katsnelson, K. S. Novoselov, A. K. Geim, cond-mat/0604323


\bibitem{Efetov} 
%Quantum dots in graphene
P.G.Silvestrov, K.B.Efetov,cond-mat/0606620 

\bibitem{Peeters} 
%Confined states and direction-dependent transmission in graphene quantum wells
J. Milton Pereira Jr., V. Mlinar, F. M. Peeters, P. Vasilopoulos,
Phys. Rev. B{\bf 74}, 045424 (2006)
%cond-mat/0606558

\bibitem{APL}
%Analysis of graphene nanoribbons as a channel material for field-effect transistors
B. Obradovic, R. Kotlyar, F. Heinz, P. Matagne, T. Rakshit, M. D. Giles, M. A. Stettler, and D. E. Nikonov
Appl. Phys. Lett. {\bf 88}, 142102 (2006)


\bibitem{Geli96} 
%Unconventional Quasiparticle Lifetime in Graphite
J. Gonz\'alez, F. Guinea, and M. A. H. Vozmediano
Phys. Rev. Lett. {\bf 77}, 3589-3592 (1996)

%\bibitem{Guinea-Confinement}
%Confinement WITH magnetic FIELD
%N. M. R. Peres, A. H. Castro-Neto F. Guinea, 
%cond-mat0603771
%Phys. Rev. B {\bf 73}, 241403 (2006)


\bibitem{Guineamagnetism}
%cond-mat/0505557
%Local defects and ferromagnetism in graphene layers
 M. A. H. Vozmediano, M. P. Lopez-Sancho, T. Stauber, F. Guinea
 Phys. Rev. B {\bf 72}, 155121 (2005)
%cond-mat/0507061 
%Coulomb Interactions and Ferromagnetism in Pure and Doped Graphene
N. M. R. Peres, F. Guinea, A. H. Castro Neto
Phys. Rev. B {\bf 72}, 174406 (2005)


\bibitem{Spin-orbit-NT} 
%Spin Splitting Induced by Spin-Orbit Interaction in Chiral Nanotubes
L. Chico, M. P. L\'opez-Sancho, and M. C. Mu\~noz
Phys. Rev. Lett. {\bf 93}, 176402 (2004)


\bibitem{Spin-orbit-dani} D. Huertas-Hernando, F. Guinea, A. Brataas,
 cond-mat/0606580 
%Spin-orbit coupling in curved graphene, fullerenes, nanotubes, and nanotube caps

\bibitem{Spin-orbit}
C.L. Kane, E.J. Mele, Phys. Rev. Lett. {\bf 95}, 226801 (2005),

\bibitem{Cortijo06} 
%Electronic properties of curved graphene sheets
 Alberto Cortijo, M. A. H. Vozmediano, cond-mat/0603717 

\bibitem{Valley-valve} 
A. Rycerz, J. Tworzydlo, C. W. J. Beenakker,
cond-mat/0608533 
%    Title: Valley filter and valley valve in graphene


\end{references}
\end{document}